\documentclass[letter,
               keeplastbox,   
               ]{jacow}

\usepackage{pdfpages,multirow,ragged2e} %
\usepackage{gensymb}
\makeatletter%
	\ifboolexpr{bool{xetex}}
	 {\renewcommand{\Gin@extensions}{.pdf,%
	                    .png,.jpg,.bmp,.pict,.tif,.psd,.mac,.sga,.tga,.gif,%
	                    .eps,.ps,%
	                    }}{}
\makeatother

\ifboolexpr{bool{xetex} or bool{luatex}} 
 {}                                      
 {\usepackage[utf8]{inputenc}}           

\usepackage[USenglish]{babel}
\usepackage{siunitx}
\usepackage{subcaption}

\ifboolexpr{bool{jacowbiblatex}}%
 {%
  \addbibresource{jacow-test.bib}
  \addbibresource{biblatex-examples.bib}
 }{}
\usepackage{tikz,xcolor,hyperref}

\definecolor{lime}{HTML}{A6CE39}
\DeclareRobustCommand{\orcidicon}{%
	\begin{tikzpicture}
	\draw[lime, fill=lime] (0,0) 
	circle [radius=0.16] 
	node[white] {{\fontfamily{qag}\selectfont \tiny ID}};
	\draw[white, fill=white] (-0.0625,0.095) 
	circle [radius=0.007];
	\end{tikzpicture}
	\hspace{-2mm}
}

\foreach \x in {A, ..., Z}{%
	\expandafter\xdef\csname orcid\x\endcsname{\noexpand\href{https://orcid.org/\csname orcidauthor\x\endcsname}{\noexpand\orcidicon}}
}

\begin{document}

\title{HOM-Damping Studies in a Multi-Cell Elliptical Superconducting RF Cavity for the Multi-Turn Energy Recovery Linac PERLE}

\newcommand{\orcidauthorA}{0000-0001-7594-5840} 

\newcommand{\orcidauthorB}{0000-0001-6346-5989} 

\newcommand{\orcidauthorC}{0000-0002-5903-8930} 

\author{C. Barbagallo\orcidA{}\textsuperscript{1}\thanks{carmelo.barbagallo@cern.ch}\thanks{Now at CERN, Geneva, Switzerland}, P. Duchesne, W. Kaabi, G. Olry, F. Zomer\textsuperscript{1}\\Laboratoire de Physique des 2 Infinis Irène Joliot-Curie (IJCLab), Orsay, France \\
		R. A. Rimmer, H. Wang, Jefferson Lab, Newport News, VA 23606, USA \\
        R. Apsimon\orcidB{}\textsuperscript{2}, S. Setiniyaz\orcidC{}\textsuperscript{2}\thanks{Now at Center for Advanced Studies of Accelerators, Jefferson Lab, Newport News, USA}, Lancaster University, LA1 4YW, UK \\
		\textsuperscript{1}also at The Paris-Saclay University, Gif-sur-Yvette, France \\	
  	\textsuperscript{2}also at Cockcroft Institute, Daresbury Laboratory, Warrington, WA4 4AD, UK}

\maketitle

\begin{abstract}
   Higher order mode (HOM) damping is a crucial issue for the next generation of high-current energy recovery linacs (ERLs). Beam-induced HOMs can store sufficient energy in the superconducting RF (SRF) cavities, giving rise to beam instabilities and increasing the heat load at cryogenic temperatures. To limit these effects, using HOM couplers on the cutoff tubes of SRF cavities becomes crucial to absorb beam-induced wakefields consisting of all cavity eigenmodes. The study presented here focuses on a 5-cell 801.58 MHz elliptical SRF cavity designed for the multi-turn energy recovery linac PERLE (Powerful Energy Recovery Linac for Experiments). Several coaxial coupler designs are analyzed and optimized to enhance the damping of monopole and dipole HOMs of the 5-cell cavity. The broadband performance of HOM damping is also confirmed by the time-domain wakefield and the frequency-domain simulations. In addition, the thermal behavior of the HOM couplers is investigated. A comparison between various HOM-damping schemes is carried out to guarantee an efficient HOM power extraction from the cavity.
\end{abstract}

\section{INTRODUCTION}

\begin{figure}[!htb]
   \centering
   \includegraphics*[width=1\columnwidth]{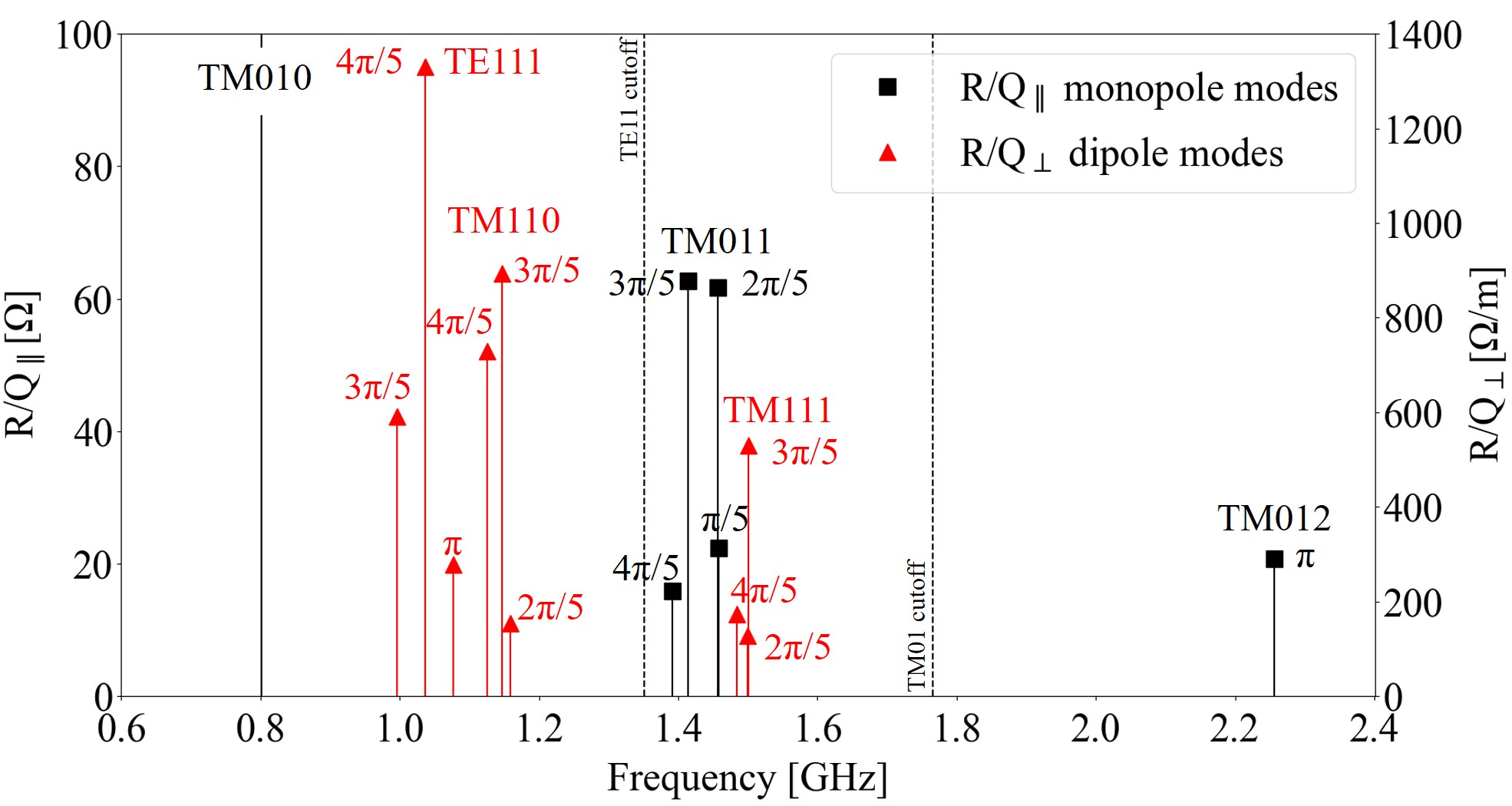}
   \caption{High \begin{math}{R/Q}\end{math} HOMs computed up to 2.4 GHz in the 5-cell bare cavity.}
   \label{fig:HOMs_5cell}
\end{figure}

PERLE~\cite{angal2018perle} is an R\&D project focused on building a three-pass ERL based on SRF technology capable of operating in the 10 MW beam power regime. In its 500 MeV configuration, PERLE will be equipped with two parallel 82 MeV superconducting linac cryomodules, each containing four 801.58 MHz 5-cell elliptical Nb cavities ($\beta$ = 1) operating with a total beam current of 120 mA in continuous-wave (CW) mode~\cite{kaabi2019perle}. In addition, coaxial couplers will need to be installed to mitigate HOMs effects. This paper proposes numerical HOM-damping studies for the 5-cell 801.58 MHz bare-cavity design proposed for PERLE by Jefferson Lab (JLab)~\cite{marhauser2018recent}. Several designs of HOM couplers are geometrically optimized in CST Studio Suite \cite{cst2021} to primarily damp high \begin{math}{R/Q}\end{math} cavity modes (Fig.~\ref{fig:HOMs_5cell}) and to meet regenerative beam breakup (BBU) and RF-heating requirements. These couplers feature probe or loop antennas designed to couple ERL optics-related cavity modes and reject the fundamental mode sufficiently. Finally, several HOM schemes for the PERLE cavity are compared, with the intention of lowering the parasitic longitudinal and transverse impedance.  

\section{Multi-pass Beam Breakup}

In high-current ERLs, a relevant effect is the regenerative multi-pass BBU, which is caused by the interaction of the recirculating electron beam with the HOMs of the accelerating cavities. The beam deflected off-axis on the first pass by the nonzero transverse electromagnetic field of dipole modes returns to the same cavity with a transverse offset on the second or higher passes~\cite{pozdeyev2005regenerative}. Similar to the transverse case, longitudinal HOMs excited by electron bunches can change the bunch energy, causing a shift in its arrival time at the cavity~\cite{song2006longitudinal}. The strength of wake fields in cavities may dramatically increase due to the feedback loop formed by the recirculating bunches. Consequently, the beam becomes unstable if the average beam current exceeds a certain threshold value.

For a single cavity containing a HOM with frequency $\omega_{\lambda}$ driving instability for a bunch in a two-pass ERL, the analytical expression for the BBU threshold current is given by~\cite{krafft1990calculating,
hoffstaetter2004beam, yunn2005expressions}: 

\begin{equation}\label{BBU current threshold 1-turn ERL} 
    I_{\mathrm{th}}=-\frac{2E}{e(R/Q)_{\lambda}Q_{\lambda}k_{\lambda}M_{mn}\sin{(\omega_{\lambda} t_{r}})},
\end{equation}

\noindent where $t_{r}$ is the recirculation time, $E$ is the energy of the beam in the recirculation arc, \begin{math}M_{mn}\end{math} is the ($m$,$n$) element of the beam transfer matrix, $e$ is the elementary charge, \begin{math}(R/Q)_{\lambda}Q_{\lambda}\end{math} is the impedance (in units of \si{\ohm}) of the considered HOM, \begin{math}Q_{\lambda}\end{math} is the quality factor, and $k_{m}=\omega_{\lambda}/c$ is the wave number with $c$ the speed of light. The Eq.~\eqref{BBU current threshold 1-turn ERL} is valid only for \begin{math}M_{mn}\sin{(\omega_{\lambda} t_{r}})<1\end{math}. When $\lambda$ denotes a dipole HOM, and \begin{math}m, n=1, 2\end{math}, the Eq.~\eqref{BBU current threshold 1-turn ERL} gives the threshold current of the transverse BBU. When $\lambda$ refers to a monopole HOM, and \begin{math}m, n=5, 6\end{math}, the Eq.~\eqref{BBU current threshold 1-turn ERL} returns the threshold current of the longitudinal BBU. In particular, the \begin{math}M_{12}\end{math} term relates the angular kick of the beam to the offset after one recirculation. The \begin{math}M_{56}\end{math} term correlates the amount of change in beam energy to the time shift of the beam.

For a multi-pass machine featuring more than two passes, the expression for the threshold current can be written as:

\begin{equation}\label{dipole BBU current threshold multi-turn ERL} 
    I_{\mathrm{th}}=-\frac{2E}{e(R/Q)_{\lambda}Q_{\lambda}k_{\lambda}\sum_{j>i=1}^{N_{c}}(E/E_{j})(M^{ij})_{mn}\sin{(\omega_{\lambda} t_{r}^{ij}})},
\end{equation}

\noindent where $E_{j}$ is the beam energy at checkpoint $j$, $(M^{ij})_{mn}$ is the ($m$,$n$) element of $M^{ij}$, which is the transfer matrix from checkpoint $i$ to checkpoint $j$, and $t_{r}^{ij}$ is the transit time from checkpoint $i$ to checkpoint $j$. In our analytical model, $N_{c}$ is the number of considered checkpoints corresponding to the linacs' exits. From Eq.~\eqref{dipole BBU current threshold multi-turn ERL}, the impedance requirement for a multi-pass machine operating at the desired current $I_{\mathrm{op}}$ can be calculated as:

\begin{equation}\label{transverse impedance multi-turn ERL} 
    \left(\frac{R}{Q}\right)_{\lambda}Q_{\lambda}\leq-\frac{2E}{ek_{\lambda}\sum_{j>i=1}^{N_{c}}(E/E_{j})(M^{ij})_{mn}\sin{(\omega_{\lambda} t_{r}^{ij}})I_{\mathrm{op}}}.
\end{equation}

From Eq.~\eqref{transverse impedance multi-turn ERL}, we can calculate the loaded quality factor $Q_{\mathrm{l}}$ value above which BBU instabilities may arise for each HOM (Fig.~\ref{fig:Qext_glob}). Since RF losses in superconducting cavities can be neglected, we assume that \begin{math}Q_{\mathrm{l}}\approx Q_{\mathrm{ext}}\end{math}, where \begin{math}Q_{\mathrm{ext}}\end{math} is the external quality factor. 

\section{HOM coupler optimization} 

\begin{figure}[!htb]
   \centering
     \begin{subfigure}[b]{0.53\textwidth}
      \includegraphics*[width=0.91\columnwidth]{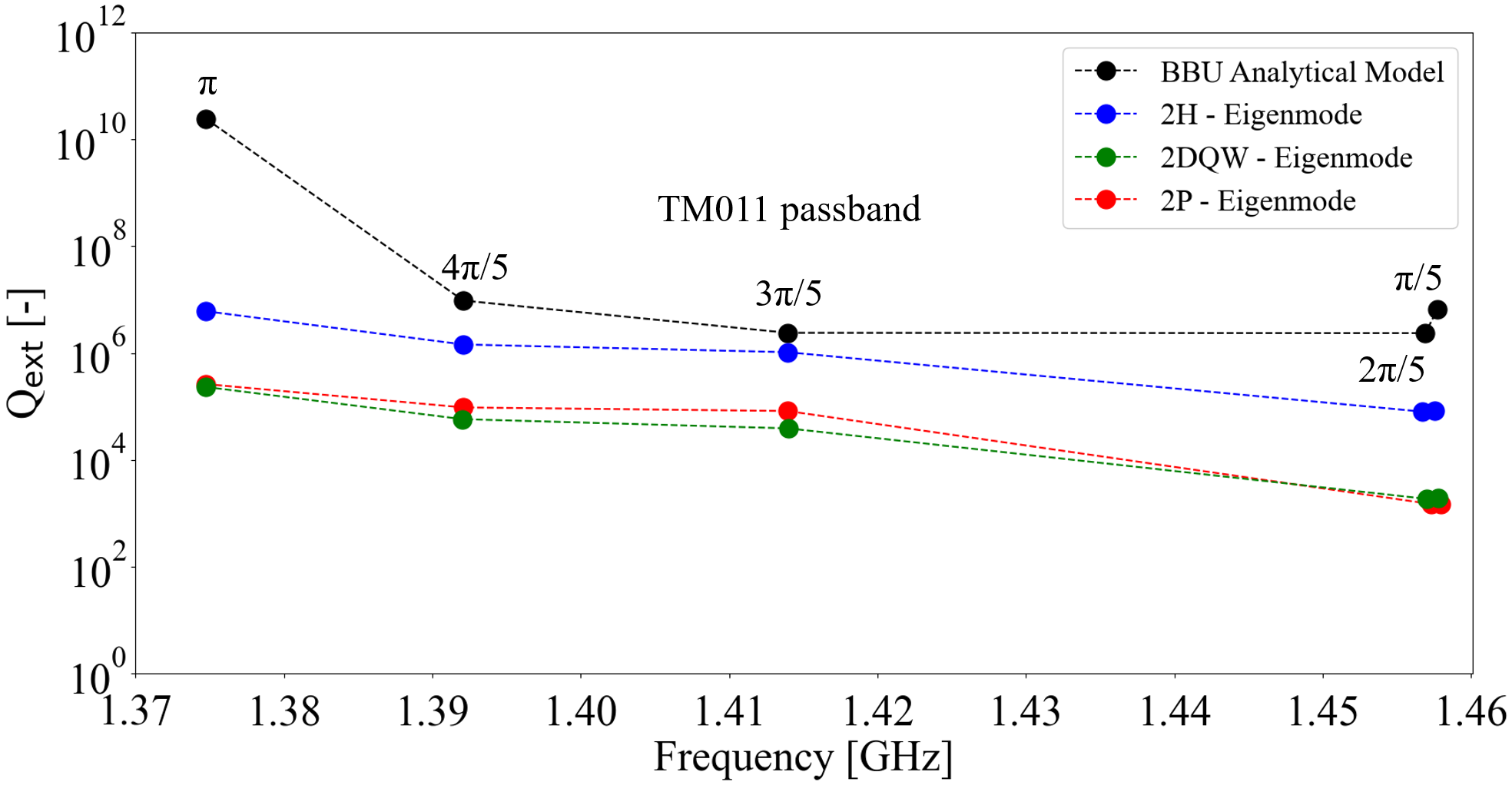}
  \caption{}
   \label{fig:Qext_mon}
   \end{subfigure}
\begin{subfigure}[b]{0.53\textwidth}
      \includegraphics*[width=0.91\columnwidth]{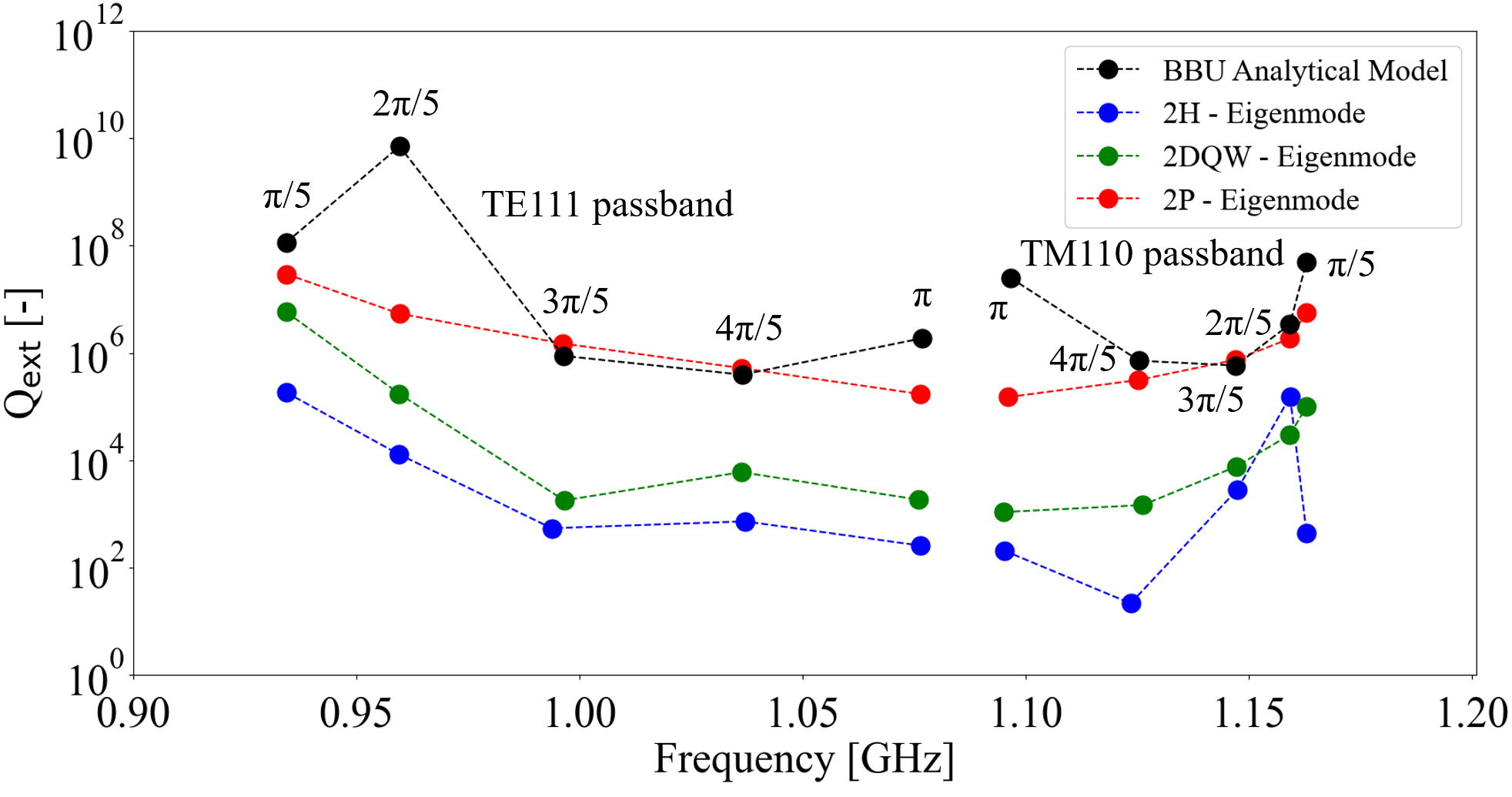}
  \caption{}
   \label{fig:Qext_dip}
   \end{subfigure}
   \caption{Simulated \begin{math} Q_{\mathrm{ext}}\end{math} of the TM011 (a), TE111 and TM110 (b) passbands for the 2H, 2DQW and 2P schemes.}
    \label{fig:Qext_glob}
 \end{figure}

The first step in finding a suitable HOM-damping scheme lies in lowering the external $Q$-value to minimize the impedance \begin{math}(R/Q)\cdot Q_{\mathrm{ext}}\end{math} of the cavity modes. Three different HOM coupler designs (Fig.~\ref{fig:HOM_damping_schemes}) were geometrically optimized using the 3D frequency domain solver of CST\cite{cst2021}: the probe-type, hook-type~\cite{Papadopoulos:1751451}, and a rescaled version of the Double-Quarter Wave (DQW) coupler~\cite{mitchell2018dqw}. A penetration depth of 20~mm into the cutoff tube is chosen for the coupling antenna of the three couplers. 
Successively, eigenmode simulations have been performed to test the performance of the following HOM-coupler schemes: two hook couplers (2H), two DQW couplers (2DQW), and two probe couplers (2P). One coupler is placed on each side of the 5-cell cavity. The probe-type and hook-type couplers are designed to mainly damp monopole and dipole modes, respectively. The DQW coupler is conceived to damp both monopole and dipole modes. Results, presented in Fig.~\ref{fig:Qext_glob}, show that the DQW coupler provides higher damping for the monopole modes than the probe coupler. The hook coupler remains the preferred solution for damping dipole HOMs. \begin{math} Q_{\mathrm{ext}}\end{math} values of monopole modes for the 2P and 2DQW schemes are lower by at least a factor 10 than the maximum allowed \begin{math} Q_{\mathrm{ext}}\end{math} values predicted by our BBU analytical model. The same conclusion is obtained for dipole modes when the 2H or 2DQW scheme is employed. Consequently, BBU's requirements for the most dangerous HOMs modes are fulfilled.     

 \section{RF-Heating analysis} 
\begin{figure}[!htb]
   \centering
      \includegraphics*[width=1\columnwidth]{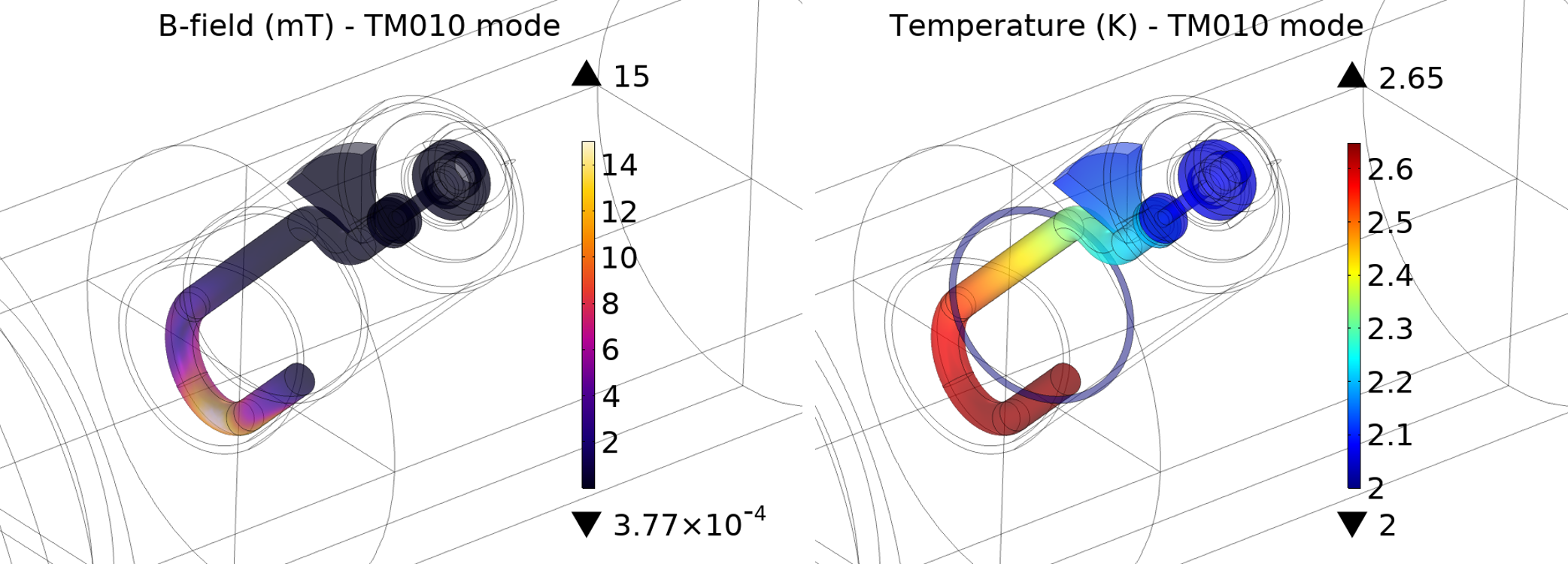}
  \caption{Magnetic field (a) and resulting temperature (b) distribution on the inner conductor of the hook coupler.}
   \label{fig:RFHeating_hook_2}
\end{figure}

This section investigates the thermal behavior of the previously analyzed HOM-damping schemes. Coupled RF-thermal simulations have been carried out in COMSOL~\cite{comsol56} to determine the average power dissipation $P_\mathrm{d}$ caused by the electromagnetic fields extracted from the cavity and the peak temperature on the HOM coupler surface. We focused our analysis on the fundamental mode only since it mainly dominates the total dissipated power. The surface resistance for the bulk Nb, related to the temperature (2 K) and RRR=300, used in the simulations is composed by \begin{math}R_{\mathrm{bcs}}\end{math} and the residual resistance \begin{math}R_{\mathrm{res}}\end{math}\cite{appleby2020science}. The value of 8.51 \si{n\ohm} is considered for the surface resistance of the cavity surface \cite{marhauser2019recent}, while 60 \si{n\ohm} is the surface resistance value assigned to the coupler antenna. Temperature is fixed at 2 K in the ring connecting the coupler port to the coupler outer conductor. The magnetic field and temperature distribution for the right-side hook coupler are shown in Fig.~\ref{fig:RFHeating_hook_2}. The results for the probe and hook couplers are summarized in Table~\ref{RFHeating_table}. The DQW coupler undergoes quenching under the studied cooling conditions.

\begin{table}[!hbt]
 \centering
  \caption{RF-heating results for the coupler inner conductor.}
   \begin{tabular}{lcccc}
     \toprule
     \textbf{}& &\textbf{$\boldsymbol{B_\mathrm{pk}}$ [mT]} & \textbf{$\boldsymbol{P_\mathrm{d}}$ [mW]} & \textbf{$\boldsymbol{T_\mathrm{max}}$ [K]} \\
      \midrule
     Hook coupler&   &                   15.01            &   1.96                 &   2.65    \\ 
     Probe coupler&  &                   29.86            &   5.02                 &   3.03    \\ 
       \bottomrule
   \end{tabular}
   \label{RFHeating_table}
\end{table}

\section{HOM-damping schemes}
\begin{figure}[!htb]
   \centering
      \includegraphics*[width=1\columnwidth]{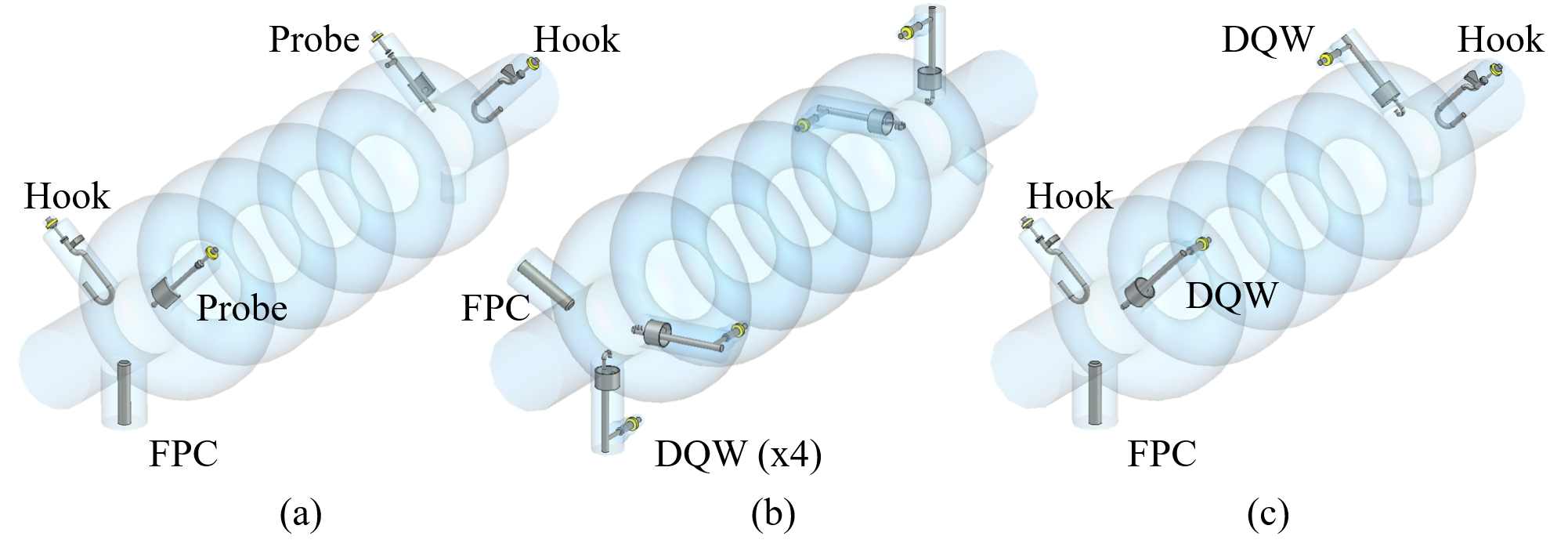}
  \caption{HOM damping schemes: a) 2H2P, b) 4DQW, c) 2H2DQW.}
   \label{fig:HOM_damping_schemes}
\end{figure}

This section includes the calculation of the broadband coupling impedance spectra using the 3D wakefield solver of CST \cite{cst2021}. In HOM-damped multi-cell SRF cavities, the broadband spectrum appears as a combination of multiple modes. For this reason, we simulated the broadband impedance spectra using a "multi-beam" wake excitation method~\cite{wang2007simulations}, which separately excites monopole, dipole, quadrupole, or even higher modes, suppressing unwanted modes. This numerical technique uses a single Gaussian bunch traveling through the cavity axis to calculate the longitudinal wake potential. Two electron bunches with an opposite charge and equal offset from the beam axis simulate the two polarizations of dipole excitation. Successively, the longitudinal and transversal impedance spectrum is obtained using a customized FFT\cite{marhauser2009enhanced} of the wake potential\cite{palumbo2003wake}.
Three different HOM-coupler schemes (Fig.~\ref{fig:HOM_damping_schemes}) are studied: two hook-type and two probe-type couplers (2H2P scheme), two hook-type and two DQW couplers (2H2DQW scheme), and four DQW couplers (4DQW scheme). HOM couplers and beam pipe ends are terminated with waveguide ports. 

\begin{figure}[!htb]
   \centering
     \begin{subfigure}[b]{0.53\textwidth}
      \includegraphics*[width=0.9\columnwidth]{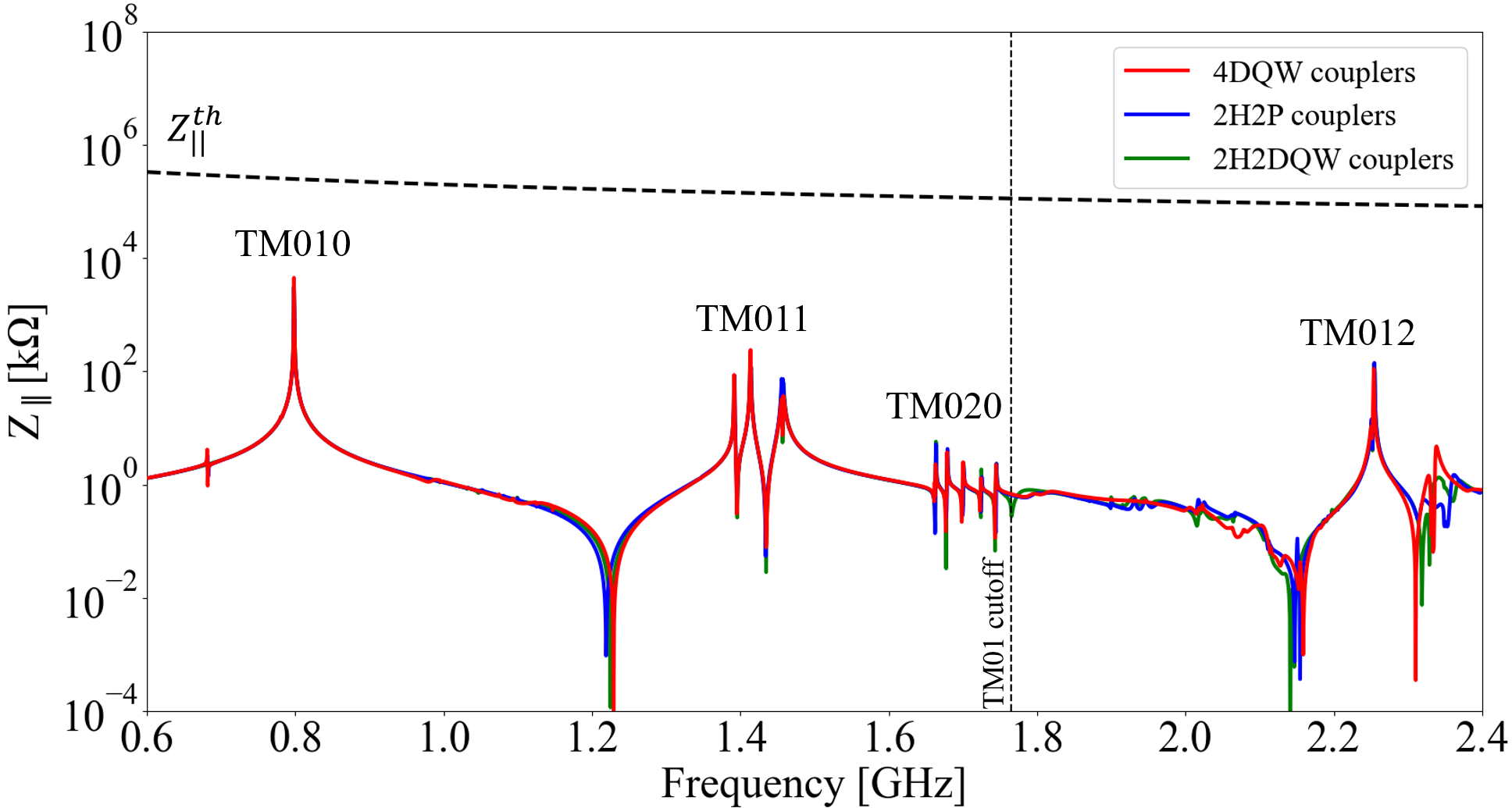}
  \caption{}
   \label{fig:ZL_damping_schemes}
   \end{subfigure}
\begin{subfigure}[b]{0.53\textwidth}
      \includegraphics*[width=0.9\columnwidth]{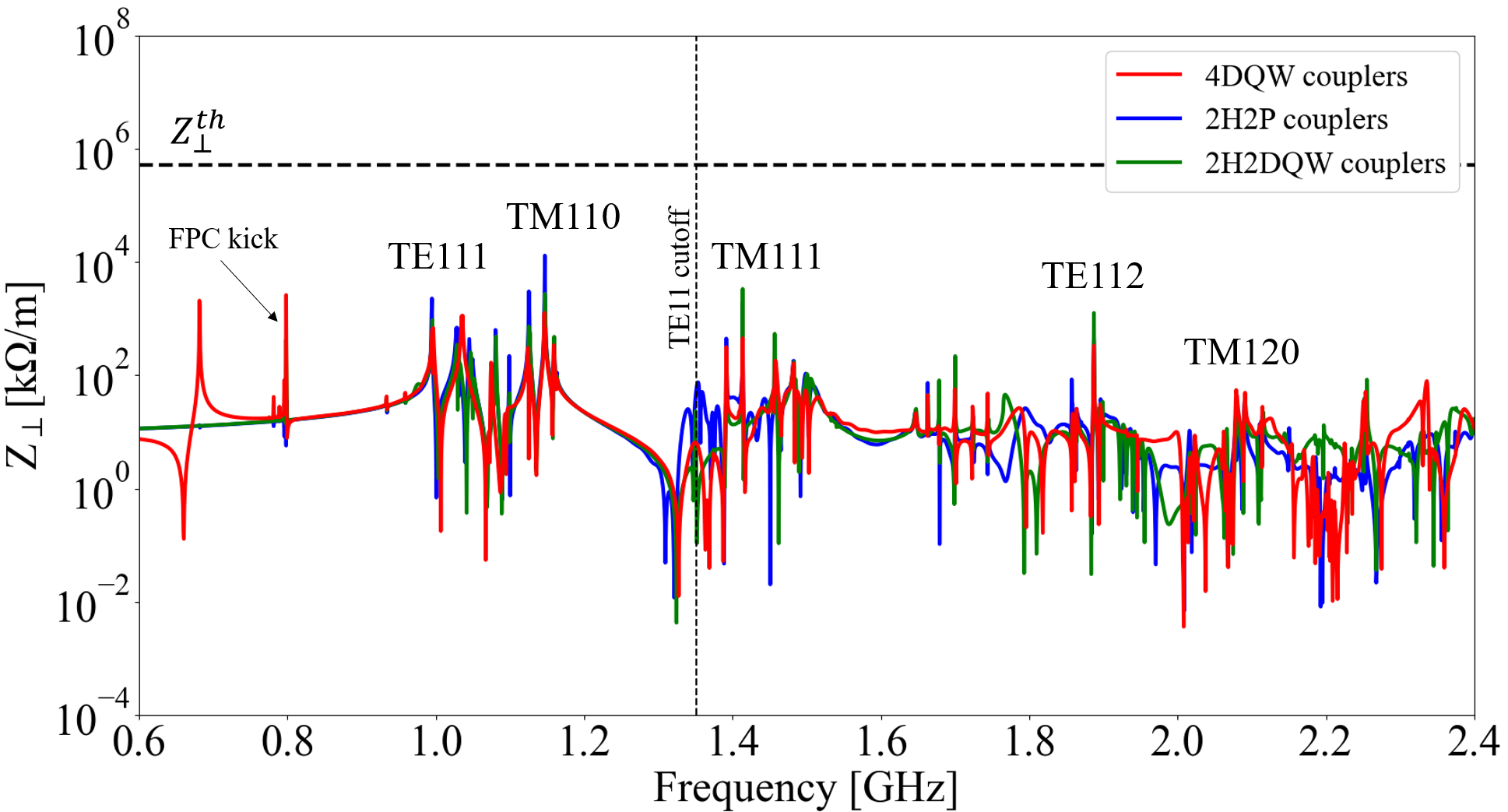}
  \caption{}
   \label{fig:ZL_damping_schemes}
   \end{subfigure}
   \caption{Longitudinal (a) and transverse (b) impedance of the cavity for the analyzed damping schemes. The wake impedance is extrapolated from a wake length of 500 m.}
   \label{fig:ZL_damping_schemes_glob}
 \end{figure}

Figure~\ref{fig:ZL_damping_schemes_glob} shows the cavity's longitudinal and transverse (for one dipole mode polarization only) impedance spectra for the analyzed damping schemes. The 4DQW scheme can lower the impedance of trapped monopole (TM011, TM020) and dipole (TE111, TM110) passbands more efficiently than the 2H2P and 2H2DQW schemes. However, different orientations of the couplers may contribute to reducing further the calculated impedance. The obtained impedance levels for the three damping schemes are lower than the longitudinal (\begin{math}Z_{\parallel}\textsuperscript{th}\end{math}) and transverse impedance threshold (\begin{math}Z_{\bot}\textsuperscript{th}\end{math}), which allows BBU's requirements to be successfully satisfied.

\section{CONCLUSION}

In this paper, a numerical investigation of different HOM-damping schemes for the PERLE cavity was carried out. An analytical multi-pass BBU model was developed to identify the maximum allowed \begin{math}Q_{\mathrm{ext}}\end{math} value and impedance threshold for each HOM. The maximum \begin{math}Q_{\mathrm{ext}}\end{math} values were compared with the external $Q$-values of three damping schemes, each containing two HOM couplers of the same type (probe, hook, DQW). RF-heating analyses were performed on the HOM couplers to calculate the maximum magnetic field, the dissipated power, and the maximum temperature on the coupler inner conductor. Finally, three different HOM damping schemes (2H2P, 4DQW, and 2H2DQW scheme) were compared in terms of impedance. Results show that the hook and the DQW coupler provide higher damping for the confined dipole and monopole HOMs, respectively. The dissipated power on the inner conductor of the analyzed couplers results in a maximum temperature lower than the Nb critical temperature, except for the DQW coupler. The DQW coupler requires an active cooling technique to maintain the superconductivity under operation. The 4DQW damping scheme showed promising results in damping both monopole and dipole HOMs below the calculated stability limits.

%

\ifboolexpr{bool{jacowbiblatex}}%
	{\printbibliography}%

\begin{thebibliography}{9} 
	
        \bibitem{angal2018perle} 
		D. Angal-Kalinin \textit{et al.},
            "PERLE. Powerful energy recovery linac for experiments. Conceptual design report",
            \textit{Journal of Physics G: Nuclear and Particle Physics}, vol. 45, no. 6, p. 065003, 2018,
\url{doi:10.1088/1361-6471/aaa171}.

        \bibitem{kaabi2019perle}
		W. Kaabi \emph{et al.},
   \textquotedblleft{PERLE: A High Power Energy Recovery Facility}\textquotedblright,
   in \emph{Proc. IPAC’19}, Melbourne, Australia, May 2019, pp. 1396--1399,
   \url{doi:10.18429/JACoW-IPAC2019-TUPGW008}. 

       \bibitem{marhauser2018recent}
		 F. Marhauser \textit{et al.},
            "Recent results on a multi-cell 802 MHz bulk Nb cavity",
            presented at the FCC week 2018, Amsterdam, Netherlands, Apr. 2018.

    \bibitem{cst2021}
		CST Studio Suite 2021,
            \url{https://www.cst.com/}.

    \bibitem{pozdeyev2005regenerative}
	 E. Pozdeyev, "Regenerative multipass beam breakup in two dimensions", \textit{Physical Review Special Topics-Accelerators and Beams}, vol. 8, no. 5, p. 054401, 2005,
    \url{doi:10.1103/PhysRevSTAB.8.054401}.

     \bibitem{song2006longitudinal}
	 C. Song \textit{et al.}, "Longitudinal BBU Threshold Current in Recirculating Linacs", \textit{Tech. rep.}, Laboratory of Elementary Particle Physics, Cornell University, Aug. 2006.

     \bibitem{krafft1990calculating}
	 G. Krafft \textit{et al.}, "Calculating beam breakup in superconducting linear accelerators", \textit{Tech. rep.}, Thomas Jefferson National Accelerator Facility (TJNAF), Newport News, VA 1990.

    \bibitem{hoffstaetter2004beam}
	 G. H.  Hoffstaetter \textit{et al.}, "Beam-breakup instability theory for energy recovery linacs", \textit{Physical Review Special Topics-Accelerators and Beams}, vol. 7, no. 5, p. 054401, 2004,
    \url{doi:10.1103/PhysRevSTAB.7.054401}.

     \bibitem{yunn2005expressions}
	 B. C. Yunn, "Expressions for the threshold current of multipass beam breakup in recirculating linacs from single cavity models", \textit{Physical Review Special Topics-Accelerators and Beams}, vol. 8, no. 10, p. 104401, 2005,
    \url{doi:10.1103/PhysRevSTAB.8.104401}.

    \bibitem{merminga2003high}
    L. Merminga \textit{et al.}, "High-current energy-recovering electron linacs", \textit{Annual Review of Nuclear and Particle Science}, vol. 53, no. 1, p. 387-429, 2003,
    \url{10.1146/annurev.nucl.53.041002.110456}.

 \bibitem{Papadopoulos:1751451}
		 P. Sotirios,
            "Higher Order Mode Couplers Optimization for the 800 MHz
            Harmonic System for HL-LHC. HL-LHC (High Luminosity - Large
             Hadron Collider)",
             CERN-STUDENTS-Note-2014-114, CERN, Geneva, Switzerland, Aug. 2014.  
                       
        \bibitem{mitchell2018dqw}
		 J. A. Mitchell \textit{et al.},
            "DQW HOM coupler design for the HL-LHC",
                       in \textit{Proc. 9th Int. Particle Accelerator Conf. (IPAC2018)}, Vancouver, BC, Canada, 2018.
                        \url{doi: 10.18429/JACoW-IPAC2018-THPAL018}. 

            \bibitem{comsol56}
		COMSOL Multiphysics 5.6,
            \url{https://www.comsol.com/}.

            \bibitem{appleby2020science}
		 R. Appleby \textit{et al.},
            "The science and technology of particle accelerators",
                       \textit{Taylor \& Francis}, 2020.

         \bibitem{marhauser2019recent}
		 F. Marhauser,
            "PERLE Cavity Design and Results and First Thoughts on HOM-Couplers",
            presented at the PERLE HOM Coupler Meeting, CERN, Geneva, Oct. 2019.

  \bibitem{wang2007simulations}
		 H. Wang \textit{et al.},
            "Simulations and measurements of a heavily HOM-damped multi-cell SRF cavity", in \textit{Proc. of PAC07}, Albuquerque, New Mexico, USA, 2007.\url{doi: 10.1109/PAC.2007.4441295}. 
            
       \bibitem{marhauser2009enhanced}
		 F. Marhauser \textit{et al.},
            "Enhanced
method for cavity impedance calculations",
                       in \textit{Proc. of PAC09}, Vancouver, BC, Canada, 2009.
                     
        \bibitem{palumbo2003wake}
		 L. Palumbo \textit{et al.},
            "Wake fields and impedance",
                       \textit{arXiv preprint physics/0309023}, 2003.

	\end{thebibliography}
	{

} 

\end{document}